\documentclass[a4paper]{article}

\usepackage{INTERSPEECH2021}
\usepackage[textwidth=18mm]{todonotes}
\usepackage{hyperref}
\usepackage{balance}
\usepackage{float}

\usepackage{verbatim}
\usepackage{soul}
\usepackage{enumitem}

\title{Exploring the Importance of F0 Trajectories for Speaker Anonymization using X-vectors and Neural Waveform Models}
\name{Ünal Ege Gaznepoglu$^{1,2}$, Nils Peters$^{1,2}$}
\address{
  $^1$University of Erlangen-Nuremberg, Germany\\
  $^2$International Audio Laboratories Erlangen, Germany}
\email{ege.gaznepoglu@fau.de, nils.peters@audiolabs-erlangen.de}

\begin{document}

\maketitle
\begin{abstract}
Voice conversion for speaker anonymization is an emerging field in speech processing research. Many state-of-the-art approaches are based on the resynthesis of the phoneme posteriorgrams (PPG), the fundamental frequency (F0) of the input signal together with modified X-vectors. Our research focuses on the role of F0 for speaker anonymization, which is an understudied area. Utilizing the VoicePrivacy Challenge 2020 framework and its datasets we developed and evaluated eight low-complexity F0 modifications prior resynthesis. We found that modifying the F0 can improve speaker anonymization by as much as 8\% with minor word-error rate degradation.
\end{abstract}
 \vspace{-1.5em}
\section{Introduction}

Developments in machine learning and increasing utilization of cloud-enabled speech interfaces has drawn attention on speech privacy. The risks associated with a cloud service gathering labeled data per speaker include realistic cloning of voices and inference of various personal information (PI). Although not required by the end-goal (e.g. speech recognition) to be achieved \cite{tomashenko_introducing_2020}, age, gender, ethnic origin, health condition, and even political and religious affiliations are among the PI buried in speech data \cite{nautsch_preserving_2019}. Moreover, both the output quality and capabilities (e.g. cross-language conversion, non-parallel conversion, zero-shot conversion) of so called DeepFake speech generators are improving while the amount of required training data reduces \cite{zhao_yi_voice_2020}.

The efforts to counter such risks led the way to the foundation of VoicePrivacy Initiative \cite{tomashenko_introducing_2020}, which through a \textit{VoicePrivacy Challenge}\footnote{\url{https://www.voiceprivacychallenge.org}} introduced common goals, baselines and (objective and subjective) metrics. Concisely, the aim is to increase the error that an automated speaker verification (ASV) system attains while keeping the automated speech recognition (ASR) performance as close to the original as possible.

In our paper we will study an underexplored area: F0 trajectories and their modifications in the context of speaker anonymization. After applying various low-complexity DSP modifications to F0, ASR and ASV performances will be evaluated. The obtained intuition could enable more complex machine learning based approaches to modify these F0 trajectories. In our paper, we first describe the baseline and the evaluation metrics inherited from the latest VoicePrivacy Challenge. Previous research on F0 that guided our own experiments will be discussed. In a first experiment we evaluate the similarity of the anonymized F0 output of the baseline system with its input.
Section \ref{sec:F0_exp} presents the follow-up experiment in which we designed and evaluated eight different F0 modification methods. Finally, we conclude with a discussion of our findings and a preview of future work.

\section{Literature Review}
\subsection{Summary of the VoicePrivacy Challenge Framework}

\subsubsection{Baseline: Speaker anonymization using X-vectors and neural waveform models}
The challenge features two baselines, one being the machine learning based approach and second being a DSP based approach \cite{patino_speaker_2020}. Challenge results yield that machine learning based proposals achieved a higher level of anonymization while retaining the intelligibility. \cite{tomashenko_voiceprivacy_2020}.  Thus, for our experiments we use the machine learning based approach as the baseline system.

\begin{figure}[H]
  \centering
  \includegraphics[width=\linewidth]{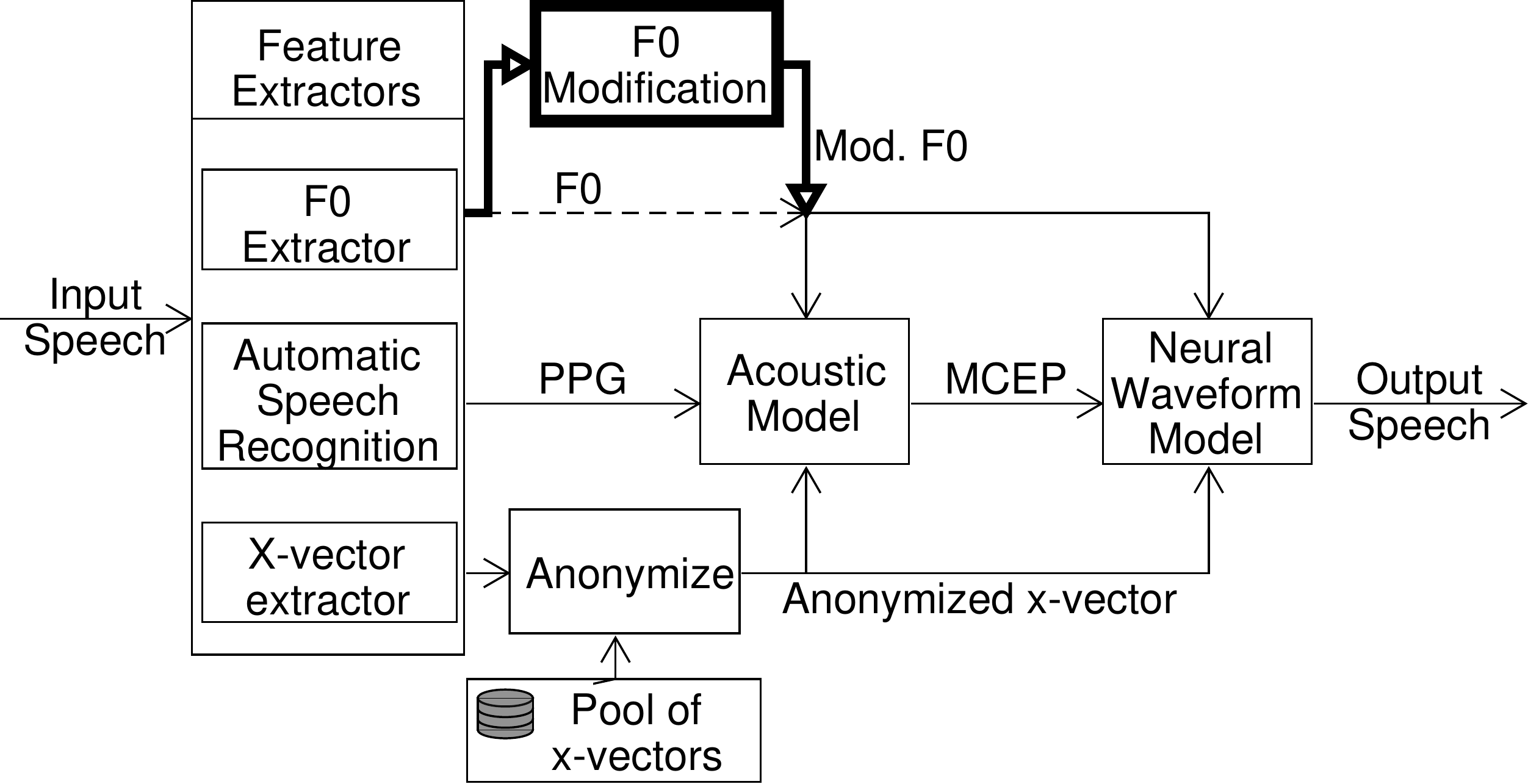}
  \caption{The VoicePrivacy 2020 Baseline 1 (blocks and arrows with thin lines and dashed line) \cite{fang_speaker_2019}, and our addition (block and arrows with thick lines).}
  \label{fig:baseline_and_our_modifications}
\end{figure}

As depicted in Figure \ref{fig:baseline_and_our_modifications}, initial feature extractors derive specialized knowledge from the input speech signal. Phonetic posteriorgrams (PPGs) are highly correlated with linguistic content. X-vectors are highly correlated with personality. 
The fundamental frequency of a speech signal is captured by the time-varying F0 trajectory. This contains knowledge not only about the personality and linguistic content, but also emotions and social context \cite{pisanski_individual_2020}. A new identity is formed according to the speaker pool, then new set of features are re-synthesized into a waveform using an acoustic model and a waveform model.

Tangled nature of F0 caused the baseline's designers to refrain from applying any modification \cite{fang_speaker_2019}. As an additional reason, they mention that ASV systems "normally use short-time spectral features rather than F0 to verify speaker identity". One of our aims is to test this claim, as we believe imposed F0 trajectories could affect the short-time spectral features.
\subsubsection{Evaluation \label{sssection:evaluation}}

The notion of privacy highly depends on the specified scenario, and the evaluation procedure has to be shaped accordingly. \\

\noindent \textbf{Attack Models:} The attacker aims to identify the speaker of an utterance. Attackers could access anonymized trial utterances, and enrolment utterances, which may or may not have been anonymized \cite{tomashenko_voiceprivacy_2020}. The anonymization system is inaccessible to the attacker. Evaluation procedure is designed accordingly.

\noindent \textbf{Evaluation Systems: }
The challenge organizers proposed objective evaluation metrics, which require ASR and ASV systems ($ASR_{eval}$, $ASV_{eval}$) to be trained. They provided the necessary system to perform the evaluation locally. 

\begin{figure}[h]
  \centering
    \includegraphics[width=\columnwidth]{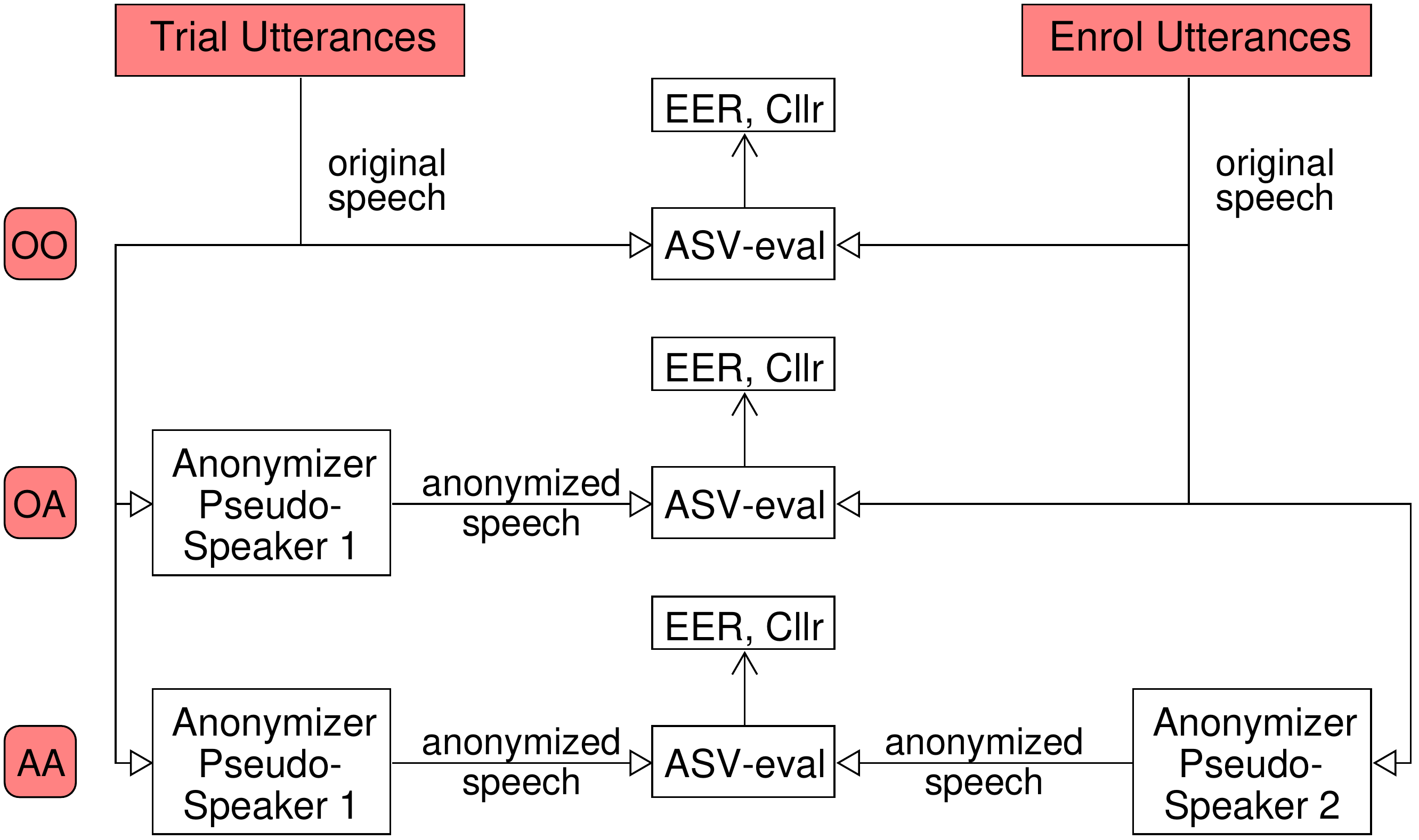}
  \caption{ASV evaluation scenarios.}
  \label{fig:asv_eval}
\end{figure}

\noindent \textbf{Evaluation of the Anonymization Performance:}
The challenge features three objective metrics, which are equal error rate (EER) and two types of the log-likelihood ratio cost function ($C_{llr}, C_{llr}^{min}$). These metrics are computed for two scenarios, 'OA' and 'AA'. 'OA' has the original signals as the enrolment data whereas 'AA' uses separately anonymized enrolment data, which should bear slightly different characteristics. Higher EER is desired in both cases, as it implies less linkability between the original and anonymized utterances. For reference, original utterances are also compared to each other ('OO'). Figure \ref{fig:asv_eval} provides a visual summary.

\noindent \textbf{Evaluation of the Intelligibility Deterioration:}
The challenge uses a single objective metric, word error rate (WER), for ASR performance evaluation. WER is computed for two different ASR systems, which employ different ('large' and 'small') language models. The anonymization process should not degrade intelligibility, thus lower WER is sought.

\subsubsection{Lessons Learned from the Challenge:}
Most promising proposals have worked on a new x-vector selection process, such as PCA-GMM based x-vector sampling \cite{turner_speaker_2021} and domain adversarial training with autoencoders \cite{espinoza-cuadros_speaker_2020}.

Reflections of the challenge organizers and contestants \cite{tomashenko_voiceprivacy_2020} provided important insight, including:
\begin{itemize}[leftmargin=*, nosep, nolistsep]
    \item 'OA' evaluations for some systems attain values exceeding 50\%, indicating full anonymization is achieved.
    \item 'AA' evaluations attain (much) lower EER scores, so same input voices even with different pseudo-speakers yield similar results. This could be a problem especially in a context where an attacker has access to the anonymizer as a black-box.
    \item WER increases significantly (e.g., from 4.15\% to 6.75\%).
    \item Cross-gender conversion has much higher WER and audible distortions are more prominent.
    \item A team supplied the system with original X-vectors and they found that both EER and WER increase \cite{champion_speaker_2020}, hinting the issues with the baseline's reconstruction capabilities. 
\end{itemize}

\noindent Based on these insights, we concluded the only unmodified feature, F0, may prove to be an interesting work direction, as all other blocks have received some attention by the contestants.

\subsection{Prior Works on F0}
\subsubsection{Speaker Identification from F0 \label{sssection:identification}}
Various statistics computed from F0 are correlated to the speaker identity. Most intuitive one is the mean of voiced F0 segments, which distinguishes male and female speakers, at least for English speakers \cite{pepiot_male_2014-1}. Given just few statistics such as log-F0 distribution skewness, average F0 rise rate, mean and variance of log-F0, it is possible to train an ASV system to distinguish speakers with EERs between 10 and 27 percent \cite{labutin_speaker_2007}. In the speaker anonymization context, this suggests that removing/altering the attributable F0 statistics is necessary.
\subsubsection{Smoothing Splines as a F0 Pre-processing Step}
A study on recognizing emotions from F0 trajectories has employed smoothing splines as a pre-processing step \cite{dellaert_recognizing_1996}. The authors reported that "such approximation of the F0 seems to contain enough information to classify the utterances according to their emotional content". Consequently, the author's findings suggest that integrating a smoothing process into the F0 parameter path preserves non-personal components and we would like to see whether it achieves some level of anonymization as well.

\subsubsection{F0 Modification in Voice Conversion Context
\label{sssection:wavelet}}
A recent work utilizes Continuous Wavelet Transform (CWT) on a parallel dataset to analyze whether any of the temporal scales could be linked to the personality \cite{sisman_wavelet_2018}. They found out that temporal scales between 20 ms and 320 ms (f: 3 Hz to 50 Hz) are significantly more correlated with the identity. Thus, a modification to these scales could help anonymizing speech.
\subsubsection{F0 Modification in Voice Anonymization Context}

Inspired by a work on voice conversion \cite{qian_f0-consistent_2020}, one of the contestants conducted further experiments, and obtained promising results after shifting and scaling the F0 trajectory according to a joint-database of F0 statistics and X-vectors \cite{champion_study_2021}. This helped especially with the results of cross-gender conversion, both the anonymization performance and the intelligibility increased. With same gender conversion, the improvement is less prominent, at a slight increase in WER. A problem with this approach is that the modification is reversible - one could extract the F0 trajectories from anonymized samples, then if an unprocessed utterance is available, they could shift-and-scale back to its statistics. Then the system in Section \ref{sssection:identification} could still be used to infer the speaker. \\

To summarize, we believe that presence of personal information in F0 trajectories and availability of various approaches to modify F0 justifies efforts on exploring their potential in the voice anonymization context. We believe that disentangling F0, altering the speaker related characteristics, possibly also perturbating with noise during the process and then re-synthesizing could increase the anonymization performance and/or better retain the intelligibility. Proposed approaches so far are target identity dependent, such as shifting and scaling. Obtaining a more comprehensive intuition by trying out target independent transforms is a coherent next-step before experimenting with machine learning based approaches.

\section{Baseline F0 Output Evaluation}\label{sec:F0_BL}
\begin{figure}[H]
  \centering
    \includegraphics[width=1\columnwidth]{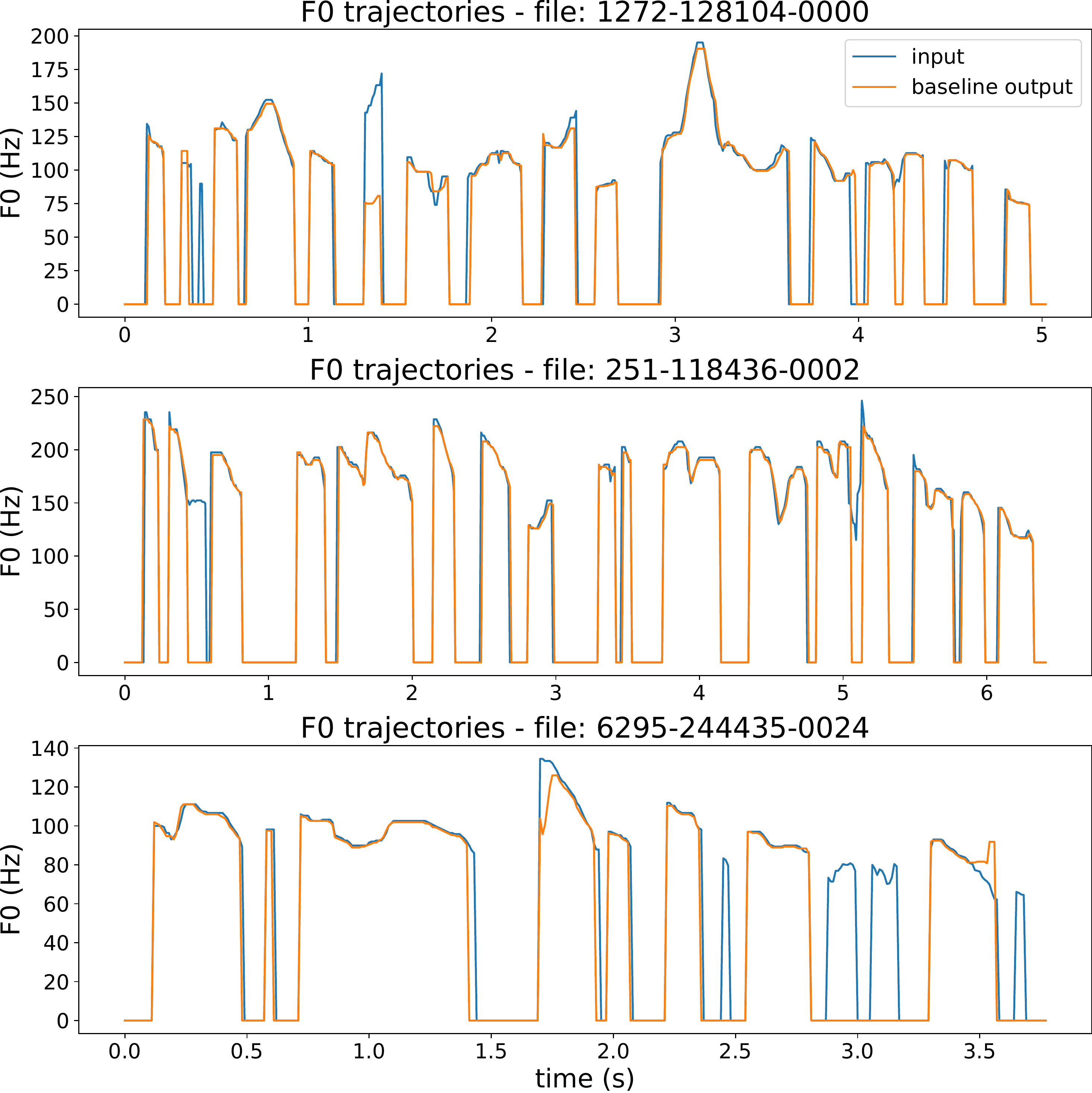}
  \caption{F0 extracted from input and baseline output. Processing delays are compensated. Scales are not normalized to maximize resolution per figure.}
  \label{fig:baseline_io_F0}
\end{figure}

Our initial experiment aims to identify how much change to F0 trajectories happen if the baseline is run, to see how much of the original F0 trajectories leak to the output and create an anonymity hazard. 
F0s are extracted using the baseline's F0 extractor (YAAPT) \cite{zahorian_spectraltemporal_2008}.
Figure \ref{fig:baseline_io_F0} shows that the output F0 trajectories follow a very similar pattern compared to the input F0, ignoring occasional disturbances. This similarity is comprehensible, as the output of neural-waveform models have consistent F0 with the input F0 \cite{wang_neural_2019} (see also Figure \ref{fig:baseline_and_our_modifications}). Evidently, the similarity of the F0 data at the output of the system indicates that the baseline architecture does not entirely eliminate the personality of the original voice. 

\section{F0 Modification Experiments}\label{sec:F0_exp}
\subsection{Description of F0 Modifiers}

\begin{figure}[!h]
  \centering
    \includegraphics[width=\columnwidth]{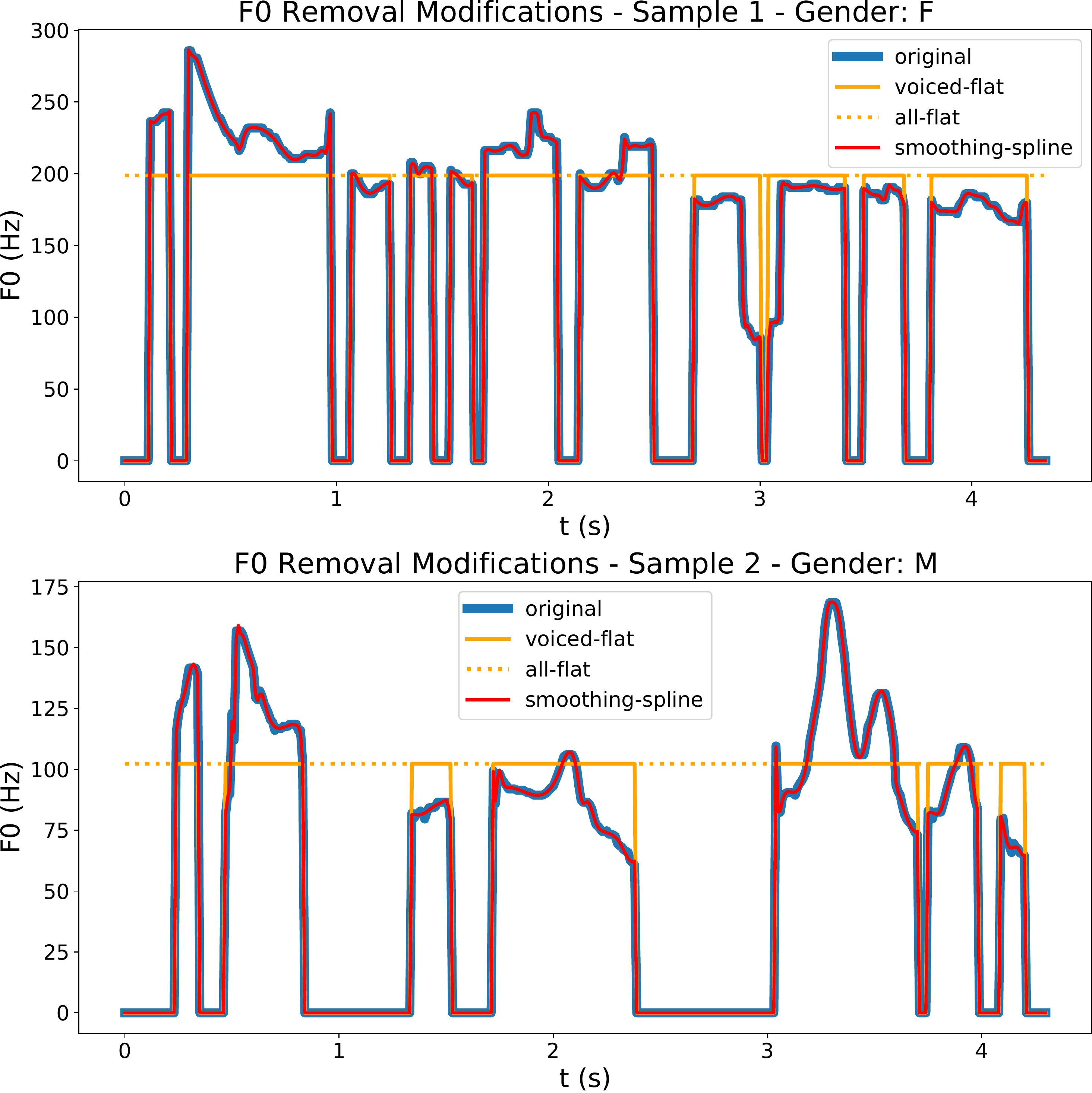}
  \caption{Proposed F0 modifications which only remove information from the original trajectories. }
  \label{fig:F0_modifications_removal}
\end{figure}

\begin{figure}[H]
  \centering
    \includegraphics[width=\columnwidth]{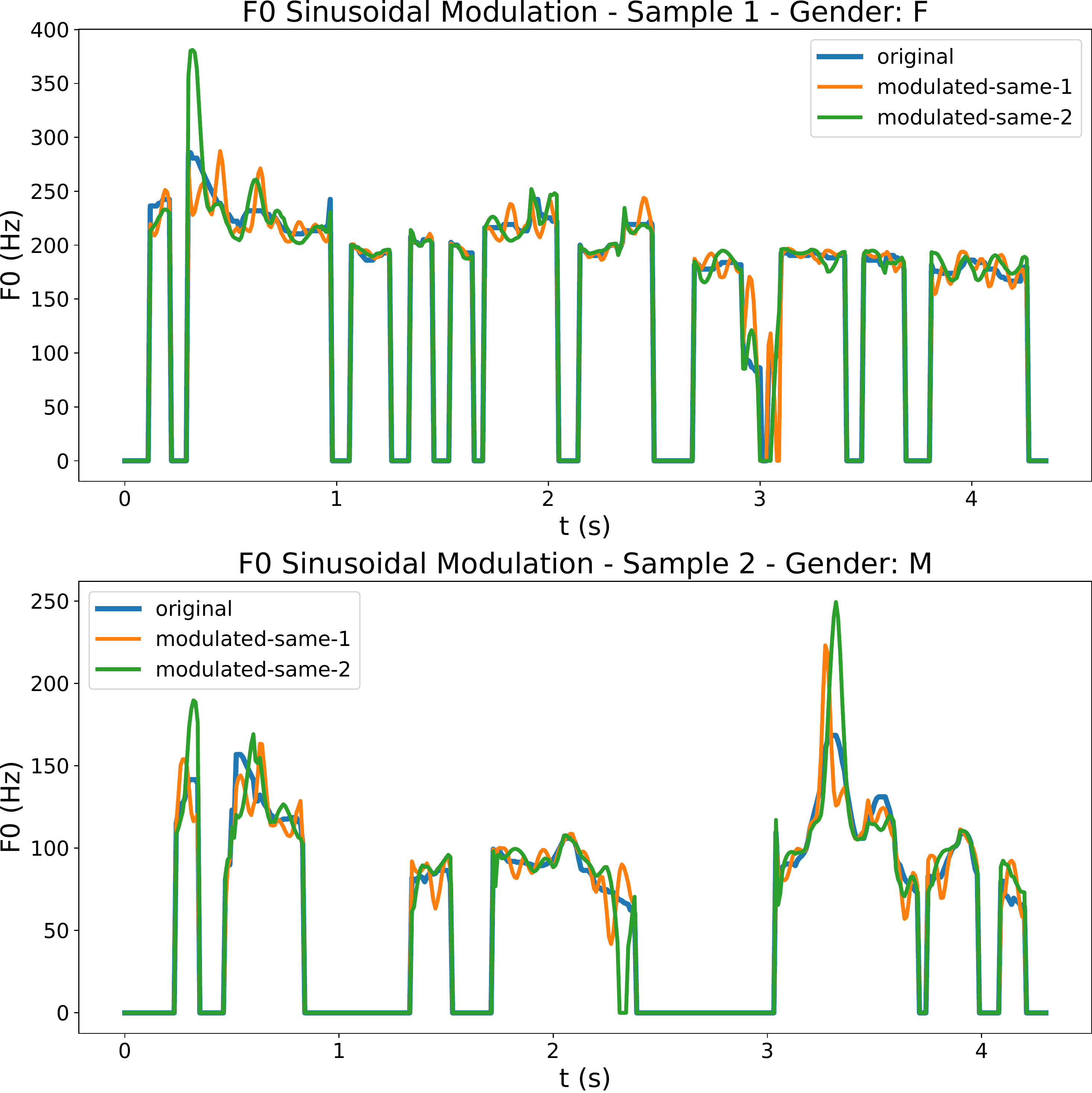}
  \caption{Proposed F0 modifications which modulate the original trajectories with quadrature sinusoidals.}
  \label{fig:F0_modifications_sinusoidal}
\end{figure}

\begin{figure}[H]
  \centering
    \includegraphics[width=\columnwidth]{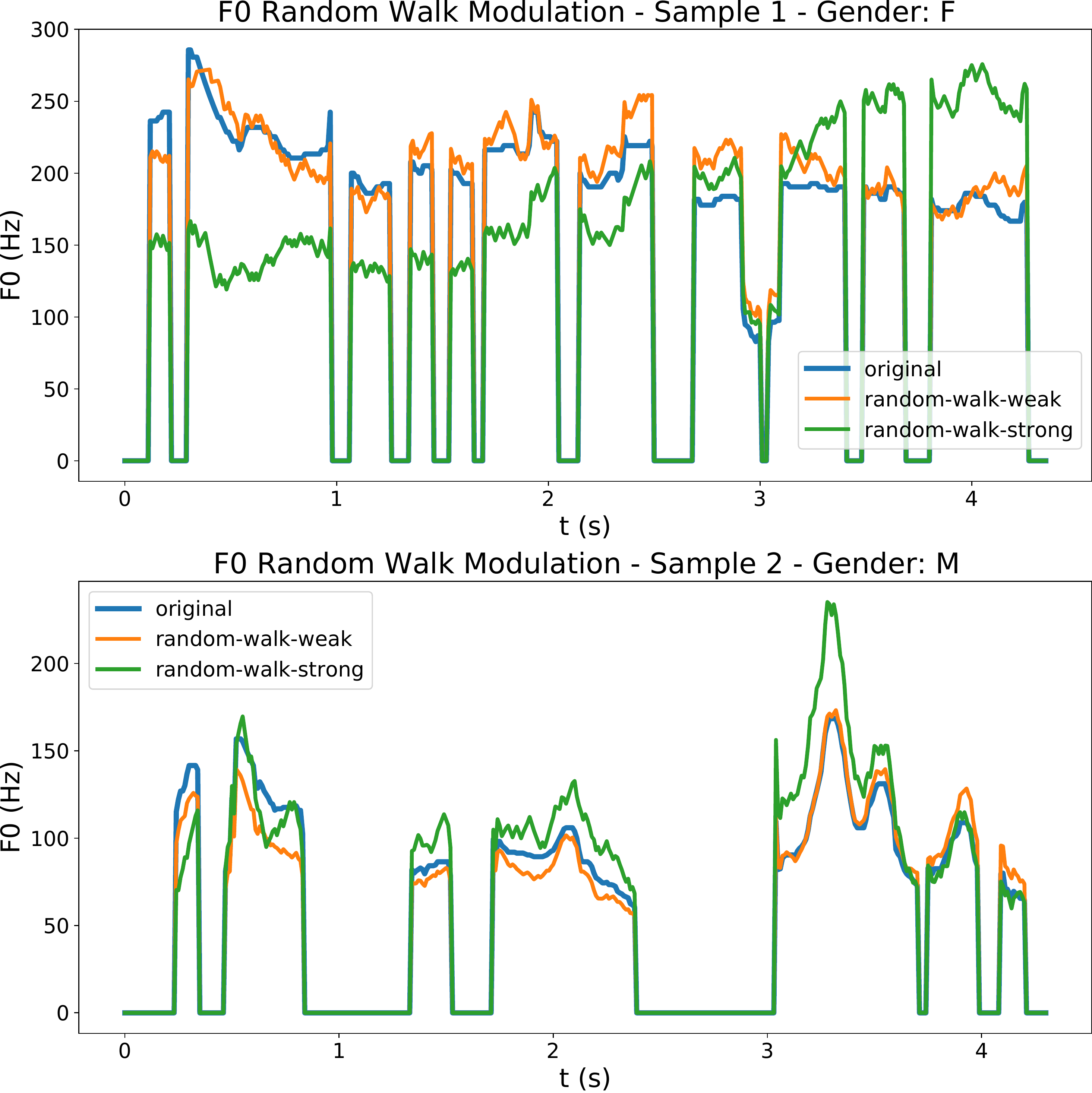}
  \caption{Proposed F0 modifications which modulate the original trajectories with uniquely generated random walk noise.}
  \label{fig:F0_modifications_random_walk}
\end{figure}

Based on the insight provided by previous works, we designed different F0 modifiers. Whereas some remove certain information from the original F0, others superpose noise or disturbances. Unvoiced segments always have their F0 set to 0, except the all-flat modification as described in Section \ref{sssection:allflat}. Also, if a modification yields impossible pitch values, (e.g., below 40 Hz, including negative values) these are set as unvoiced. All other features are fixed for all experiments. Figures \ref{fig:F0_modifications_removal}, \ref{fig:F0_modifications_sinusoidal} and \ref{fig:F0_modifications_random_walk} illustrate the modifications on two sample recordings. 

As the starting point of our experiments, we have used the default baseline settings. The new X-vector is decided on the speaker level (i.e., all excerpts from a speaker to be anonymized are analyzed, then a common pseudo-identity is used for all). $N=200$ same gender X-vectors with farthest PLDA distances to the original are collected, then $N'=100$ elements are picked randomly and their average gives the new X-vector.

\subsubsection{Flattening the Voiced Segments (reference: voiced-flat)}
For each recording, F0 mean of the voiced segments are computed, then pitch value at all voiced intervals are set to that.

\subsubsection{Flattening the Entire trajectory (reference: all-flat) \label{sssection:allflat}}
For each recording, F0 mean of the voiced segments are computed, then pitch value at everywhere are set to that. Unvoiced segments do not have their F0 forced to 0.

\subsubsection{Smoothing the F0 trajectories with Smoothing Splines}

For each recording, a smoothing cubic spline with default parameters as implemented in \textit{scipy.interpolate.UnivariateSpline} is fit to the F0 trajectory using only the voiced samples.

\subsubsection{Modulating the F0 (reference: modulated-same-*) \label{sssection:modulation}}
For each recording, F0 trajectories are modulated with two pre-determined quadrature sinusoidal waves. The wave frequencies are $ f_{1,2} = [5, 11] $ for \textit{same-1} and $ f_{1,2} = [3, 7]$ for \textit{same-2} which are picked such that they, their sums and differences are within the scales indicated in Section \ref{sssection:wavelet}, and they are prime numbers (for maximum orthogonality). Afterwards unvoiced parts are forced to zero. The mathematical equation governing the modulation is given below, where $\bar{f}$ denotes mean of the voiced parts, $ f_c(t)$ denotes a zero-centered F0 trajectory per $\bar{f}$, and $c_{1,2}(t) = sin(2\pi f_i t + \frac{(i-1)\pi}{2})$ denote "carrier" waves:
\begin{equation}
f_{out}(t) = \bar{f} + f_c(t) \frac{4 + 2 c_1(t) + 2 c_2(t) +  c_1 (t) c_2 (t)}{4}
\label{eq:modulation}
\end{equation}

\subsubsection{Modulating the F0 Trajectory with Different Frequency Pairs (reference: modulated-different)}

For each recording, F0s are modulated with two pre-determined quadrature sinusoidal waves. The frequencies are $ f = [5, 11] $ for enrollment datasets, and $ f = [3, 7] $ for trial datasets. Both the modulation methodology and the frequency selection adhere to the rules in Section \ref{sssection:modulation}.

\subsubsection{Random Walk Modulation (reference: random-walk-*)}

For each recording, a random walk signal is generated. In order to put a reasonable constrain to the interval that the F0 can fluctuate within, we applied a normalization such that extreme values of the noise become $ [-\frac{1}{2}, \frac{1}{2}] $. Then this normalized signal, denoted with $r(t)$, is applied with the following equation, with $M=1$ for 'weak' and $M=2$ for the 'strong' experiment, signifying the strength:

\begin{equation}
f_{out}(t) =  f(t) \frac{2 + M r(t)}{2}
\label{eq:random-walk}
\end{equation}

\subsection{Evaluation and Results}\label{sec:eval}

We are inheriting the datasets and the evaluation framework from VoicePrivacy Challenge as explained in Section \ref{sssection:evaluation} and computed the EER as the ASV metric (see Table \ref{tab:results_asv}) and the WER as the ASR metric (see Table \ref{tab:results_asr}). We have used the entire libri-dev, libri-test, vctk-dev and vctk-test datasets for our computations which in total contain 13982 utterances. Verbose information regarding the dataset compositions could be obtained from \cite{tomashenko_voiceprivacy_nodate}. 
The EER scores of the ASV evaluation are averages of male and female evaluations on Libri- subsets. We also appended corresponding scores from various other works \cite[Table 1,3]{turner_speaker_2021}, \cite[Table 2,5]{espinoza-cuadros_speaker_2020} using the same framework to compare the relative effectiveness of various improvements over the same baseline. Authors of \cite{champion_study_2021} also kindly shared their verbose score table with us.

For the WER scores of the ASR evaluation, we display the results of the large language model. The scores for the small language model (not shown here) follow a similar trend.
The ASR evaluation is designed to be run only on the trial dataset. Experiments modulated-diff and modulated-same-2 share the same trial dataset, thus the same WER is attained.

\begin{table}[H]
\footnotesize
\centering
\caption{ASV evaluation results for our modifications, best are emphasized}
\label{tab:results_asv}
\begin{tabular}{l|cc|cc}
\hline \hline
\textbf{}           & \multicolumn{2}{c|}{\textbf{\begin{tabular}[c]{@{}c@{}}ASV - LibriDev\\ EER (\%)\\ (higher better)\end{tabular}}} & \multicolumn{2}{c}{\textbf{\begin{tabular}[c]{@{}c@{}}ASV - LibriTest\\ EER (\%)\\ (higher better)\end{tabular}}} \\ \hline
\textbf{Method}     & \multicolumn{1}{c|}{\textbf{OA}}                                 & \textbf{AA}                                    & \multicolumn{1}{c|}{\textbf{OA}}                                 & \textbf{AA}                                    \\ \hline
Raw Data            & 4.95                                                             & N/A                                            & 4.39                                                             & N/A                                            \\
Baseline            & 54.02                                                            & 35.41                                          & 50.32                                                            & 33.63                                          \\
voiced-flat         & \textbf{54.73}                                                            & 33.24                                          & \textbf{51.15}                                                   & 32.35                                          \\
all-flat            & 54.61                                                   & 30.98                                          & 50.83                                                            & 30.17                                          \\
smoothing-spline    & 54.25                                                            & 35.40                                          & 50.23                                                            & 33.83                                          \\ 
modulated-same-1      & 54.15                                                            & 35.33                                          & 50.14                                                            & 33.69                                          \\
modulated-same-2    & 53.69                                                               & 35.83                                             & 50.36                                                               & 34.19                                             \\
modulated-different & 53.69                                                               & 35.62                                             & 50.36                                                               & 34.21                                             \\
random-walk-weak          & 54.56                                                            & 36.56                                          & 50.79                                                            & 34.63                                          \\
random-walk-strong        & 54.59                                                            & \textbf{37.21}                                 & 50.08                                                            & \textbf{36.31}                                 \\ \hline
f0-shift-and-scale \cite{champion_study_2021} & 55.14 & 36.61 & 50.78 & 38.68 \\
x-vector-gmm-pca  \cite{turner_speaker_2021} & 46.75 & 39.15 & 45.70 & 39.45 \\
x-vector-domain-adv  \cite{espinoza-cuadros_speaker_2020} & 53.95 & 35.48 & 49.69 & 34.44 \\ \hline \hline

\end{tabular}
\end{table}

\begin{table}[H] 
\caption{ASR evaluation results, which are computed only on trial subsets.}
\footnotesize
\centering
\label{tab:results_asr}
\begin{tabular}{l|cc|cc}
\hline \hline
                      & \multicolumn{2}{c|}{\textbf{\begin{tabular}[c]{@{}c@{}}ASR - WER (\%)\\ (lower better)\\ VCTK\end{tabular}}} & \multicolumn{2}{c}{\textbf{\begin{tabular}[c]{@{}c@{}}ASR - WER (\%)\\ (lower better)\\ Libri \end{tabular}}} \\ \hline
\textbf{Method}       & \multicolumn{1}{c|}{\textbf{Dev}}                              & \textbf{Test}                               & \multicolumn{1}{c|}{\textbf{Dev}}                               & \textbf{Test}                              \\ \hline
Raw Data              & 10.79                                                          & 12.79                                       & 3.83                                                            & 4.15                                       \\
Baseline              & 15.38                                                          & \textbf{15.22}                                       & 6.32                                                            & \textbf{6.71}                                       \\
voiced-flat           & 15.69                                                          & 15.48                                       & 6.42                                                            & 6.93                                       \\
all-flat              & 16.22                                                          & 15.80                                       & 6.81                                                            & 7.25                                       \\
smoothing-spline      & \textbf{15.34}                                                 & 15.25                              & \textbf{6.29}                                                   & 6.75                              \\
modulated-same-1      & 15.74                                                          & 15.57                                       & 6.65                                                            & 6.97                                       \\
modulated-same-2       & 15.55                                                             &      15.36                                     & 6.57                                                              & 6.93                                         \\
modulated-diff        & 15.55                                                          & 15.36                                      & 6.57                                                           & 6.93                                         \\
random-walk-weak            & 15.54                                                          & 15.37                                       & 6.38                                                            & 6.89                                       \\
random-walk-strong          & 15.96                                                          & 15.88                                       & 6.74                                                            & 7.12                                       \\ \hline
f0-shift-and-scale \cite{champion_study_2021} & 15.50 & 15.29 & 6.43 & 6.92 \\
x-vector-gmm-pca \cite{turner_speaker_2021} & 15.56 & 15.63 & 6.75 & 7.26 \\
x-vector-domain-adv \cite{espinoza-cuadros_speaker_2020} & 15.20 & 15.16 & 6.75 & 6.74 \\ \hline \hline
\end{tabular}
\end{table}

From Table \ref{tab:results_asv}, we see that the experiment \textit{voiced-flat} attained the highest 'OA' scores whereas \textit{random-walk-strong} attained the highest 'AA' scores, both for development and test datasets. From Table \ref{tab:results_asr}, we see that the experiment \textit{smoothing-spline} obtained WER that is even lower than the baseline itself on the training datasets, and on the test datasets it was a close second to the baseline. Listening samples are made publicly available on \footnote{\url{https://www.audiolabs-erlangen.de/fau/professor/peters/publications/MLSLP2021}}.

\section{Discussion and Future Work}

\subsection{Baseline F0 Output Evaluation}

Mostly, the F0 extracted from the baseline output was consistent with the input F0. We noticed that some originally voiced segments are identified as unvoiced at the output. We guess that this is due to the severe local distortions, hindering the significance of the pitch, thus causing the unvoiced decision at the F0 extractor. We believe that further characterization of the F0 artifacts could help us build a more robust anonymizer.

\subsection{F0 Modification Experiments}
Flattened F0 trajectories increase the 'OA' scores, yet decrease the 'AA' scores. This is consistent because originals and anonymized data have less in common, at the cost of losing the natural intra-cluster variance, as now they bear the same monotonicity. So it becomes easier to link the different anonymizations of the same original speech. The more the data gets removed, the more the WER increases. Although the WER increase is not significant, in our opinion using these systems in a real application would not yield significant benefits.

We observed that processing the F0 with smoothing splines did not yield any adverse effects, though it did not achieve any anonymization either.

In contrast, altering the F0 trajectories either with certain sinusoids, or even better with random walk noise as we have generated is a promising addition. A nice attribute is that such a modification could not be reversed, because the signal that we modulate is not properly band-limited and as a result we introduce some aliasing. We experimented with different sets of frequencies, and ASR and ASV performance changed differently according to the frequency set picked. Comparing the scores of experiments \textit{modulated-same-*} and \textit{modulated-different}, we observed that having different frequencies for different speakers did not create a significant benefit. The sinusoid modulation caused easily noticable vibrato artifacts at the output. Random walk noise on the other hand, did not create any noticeable artifacts as long as the amplitude was kept small. Increasing the strength causes audible unnatural behavior, such as inappropriately high or low F0 for given gender at localized spots or a perception of unnatural intonation.

As all works have obtained 'OA' scores that are around 50\% (corresponding to full anonymization), we will focus on the 'AA' scores. Our experiment \textit{random-walk-weak} attained similar performance in WER and training-set EER with compared shift-and-scale approach, and as a plus the reversibility handicap is remedied. Test-set EER though, is much higher for the shift-and-scale approach. Our experiment \textit{random-walk-strong} has comparable EER performance with respect to the proposed X-vector modifications. An exception to that is 'AA' scores of the domain-adversarial autoencoders approach \cite{espinoza-cuadros_speaker_2020} has a slightly lower 'AA' score, and a smaller test-set WER.

Being able to improve the anonymization performance while essentially preserving the speech intelligibility suggests that ignoring F0 is a sub-optimum approach. The community should strive towards effective F0 modification techniques. 

\subsection{Future Work}
Our findings indicate that applying low-complexity modifications to the F0 trajectories can increase anonymization by as much as 8\% (which is attained by applying \textit{random-walk-strong} and in terms of relative gain with respect to the baseline), and further experiments on F0 trajectories are worthwhile.

Based on these promising findings, a variety of further inquiries based on our research are possible. Our plans include further characterizing the F0 disturbances caused by the baseline, and to design a system which could manipulate the F0 trajectories to minimize such disturbances or other unnatural artifacts, while still being beneficial for the anonymization. 

Also, future work will continue to find optimum F0 manipulation methods to maximize the ASV's EER while avoiding an increase of the ASR's WER. 
For instance, by combining the random walk with a smoothing spline methodology as a post-processing step we believe some of the unnatural behavior introduced by our modifications may be tamed, while allowing for the desired anonymization benefit.

We would also like to continue the system evaluation with speech corpora that feature other demographics, such as elderly or children voices, to gain better understanding of the anonymization performance across the larger population. 

\section{Conclusion}

In this work we have explored various low-complexity DSP modifications to the F0 trajectories to understand their effects on the EER and WER scores. We first analyzed how the F0 trajectories change after signals are processed with the baseline. We found out that ignoring few characterizable changes, the trends look similar, which suggests identity leakage. Then eight F0 modifications were designed, implemented, and tested on F0 trajectories before re-synthesizing, including flattening, spline smoothing, sinusoidal modulation, and random walk modulation. Modulating the F0 trajectories with uniquely generated random walk noise yielded anonymization improvements of about 8\% with minor WER degredation. Since F0 manipulation has not yet been considered in the winning voice privacy systems \cite{tomashenko_voiceprivacy_2020}, we believe our method is complementary and thus, can potentially improve state-of-the-art voice privacy systems. 

\balance
\tiny
\bibliographystyle{IEEEtran}
\bibliography{mybib}

\end{document}